\documentclass[aps,prl,twocolumn,nofootinbib,longbibliography,10pt]{revtex4-2}

\usepackage[utf8]{inputenc}
\usepackage[T1]{fontenc}
\usepackage[english]{babel}

\usepackage{times}
\usepackage[
    per-mode=symbol,
    separate-uncertainty=false,
    multi-part-units=single,
    retain-zero-uncertainty=true
    ]{siunitx}
\usepackage{physics}
\usepackage{graphicx}
\usepackage{stmaryrd}
\usepackage{amssymb}
\usepackage{amsmath}
\usepackage{dsfont}

\usepackage{hyperref}
\hypersetup{
colorlinks=true, 
linkcolor=[RGB]{45,48,146}, 
citecolor=[RGB]{45,48,146}, 
filecolor=[RGB]{45,48,146}, 
urlcolor=[RGB]{45,48,146}} 
\usepackage{orcidlink}

\newcommand{\Fig}{Fig.\@ }

\newcommand{\mean}[1]{\langle #1 \rangle}

\begin{document}

\title{
Efficient entanglement of three remote single-atom quantum-network nodes
}

\author{Matthias Seubert$^*$\orcidlink{0009-0003-9777-8256}}
\author{Leonardo Ruscio$^*$\orcidlink{0000-0001-5353-0863}}
\author{Tobias Frank$^*$}
\author{Philip Thomas\orcidlink{0000-0002-3378-7976}}
\author{Maya Büki\orcidlink{0000-0002-5692-8284}}
\author{Gianvito Chiarella\orcidlink{0009-0008-2473-6027}}
\author{Pau Farrera\orcidlink{0000-0002-1264-0342}}
\author{Olivier Morin\orcidlink{0000-0002-0327-5646}}
\author{Gerhard Rempe\orcidlink{0000-0003-2770-0819}}
\affiliation{Max-Planck-Institut f\"{u}r Quantenoptik, Hans-Kopfermann-Strasse 1, 85748 Garching, Germany}

\def\thefootnote{*}\footnotetext{These authors contributed equally to this work}

\begin{abstract}

\noindent Entanglement distributed over a set of individually addressable qubit nodes is the enabling resource for a plethora of applications ranging from tests of quantum physics~\cite{Greenberger1989, Mermin1990} to secure and modular quantum information networks~\cite{Kimble2008, Wehner2018, Awschalom2021, Ramya2025}. Entanglement between two memory qubits has been realized on various platforms~\cite{Wei2022}, but extension to more nodes remains rare and formidably challenging~\cite{Jing2019, Pompili2021}. The principal bottleneck is the efficiency of the light-matter interfaces connecting the qubit nodes to their communication channels. Here, we efficiently generate, distribute and store a three-qubit entangled state across three independent laboratories containing single atoms coupled to optical resonators. We sequentially entangle the atoms pairwise, two by heralded photonic entanglement swapping and two by heralded state transfer. We reach a three-qubit entanglement fidelity of \SI{77(1)}{\percent} and an entanglement lifetime above \SI{200}{\micro\second}. The observed qubit correlations violate Mermin's inequality while closing the detection loophole. Our three-qubit entanglement-generation efficiency is \SI{0.16}{\percent}. This unprecedented efficiency of our scheme establishes a clear route towards multi-node quantum networks.
\end{abstract}

\maketitle

\section{Introduction}

\noindent Establishing a quantum internet and harvesting the revolutionary applications it promises, ranging from secure communication over enhanced sensing to distributed quantum computing, is one of the grand visions in the research field of quantum information~\cite{Kimble2008, Wehner2018, Awschalom2021}. The conceived network builds on entanglement between a set of distant quantum memories as a resource. Towards this goal, elementary entanglement links between memory nodes have been realised using various platforms and entanglement protocols~\cite{Wei2022}. Examples include ions~\cite{Moehring2007,Stephenson2020,Kucera2024}, neutral atoms~\cite{Hofmann2012}, single atoms in optical cavities~\cite{Ritter2012,Krutyanskiy2023}, atomic clouds~\cite{Matsukevich2006,Chou2007,Pu2021,Liu2024}, quantum dots~\cite{Delteil2016}, colour centers~\cite{Hensen2015,Knaut2024} and other solid state systems~\cite{Lago-Rivera2021}, to name a few. However, these implementations featured bipartite entanglement, and so far only few experiments explored more than two nodes with matter qubits~\cite{Jing2019, Pompili2021}.

The main bottleneck for the realization of a multi-node quantum network is the efficiency of the light-matter interfaces that connect the nodes to the communication channels. A small efficiency results in a low probability of entanglement generation per attempt and thus a success rate that decreases exponentially with increasing number of nodes~\cite{Jing2019}. A possible workaround is to increase the rate of attempts, but this does not resolve the scalability issue. Another approach is to follow a heralded repeat-until-success strategy. This improves scalability but translates into a heavier demand on the coherence time of the nodes and requires additional buffer memories~\cite{Pompili2021}. For all these reasons, efficient generation of light-matter entanglement remains one of the biggest levers for achieving practical and scalable network entanglement beyond two nodes.

\begin{figure}
    \centering
    \includegraphics[width=\linewidth]{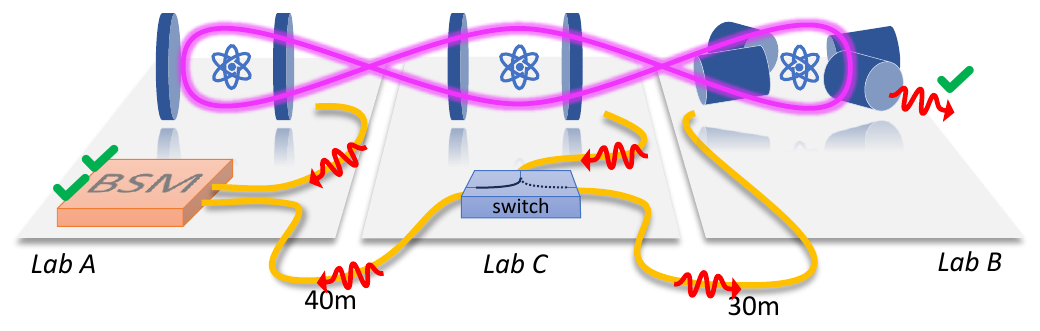}
    \caption{\textbf{Three-partite entanglement between atoms in optical cavities.} The two links between the central node C and the end nodes A and B are entangled sequentially. In a first step, Lab C generates a photon entangled with its atom and routes it through a switch to Lab A. There, an identically produced entangled atom–photon pair allows for a Bell-state measurement on the two photons, swapping the entanglement onto the two atoms. In the second step, Lab C emits another photon that is entangled with its atom. This photon is guided to Lab B and stored in a heralded memory. The green ticks indicate photon detections, which herald a successful atomic entanglement distribution. As a result a Greenberger–Horne–Zeilinger state shared by the three remote atoms is created.}
    \label{fig:setup}
\end{figure}

Here we build on the recent progress in the efficient generation of multi-photon graph states emitted from an individual atom in an optical cavity~\cite{Thomas2022}. This work has shown that it is possible to produce a stream of photons with various entanglement topologies such as a Greenberger-Horne-Zeilinger (GHZ) or cluster state. Now we employ these photons to produce a long-lived GHZ state distributed over a network with three single-atom nodes in three otherwise independent laboratories. To achieve this (see \Fig \ref{fig:setup}), we sequentially entangle two links sharing a common central node. As the two end nodes use different optical cavity architectures, two distinct entanglement techniques are employed: one uses heralded entanglement swapping via a linear-optics Bell-state measurement (BSM), the other utilises a heralded memory~\cite{Brekenfeld2020}. We first benchmark the multipartite entanglement using the Mermin inequality and observe the violation of the inequality while closing the detection loophole. We further characterize the state through the three-qubit entanglement fidelity of \SI{77(1)}{\percent} and observed an entanglement lifetime of over \SI{200}{\micro\second}, achieved without the use of spin echo or dynamical decoupling techniques. This corresponds to a fibre network communication distance of \SI{40}{\kilo\metre}. Thanks to an entanglement efficiency of \SI{0.16}{\percent} per attempt, and excluding all technical limitations not inherent to the quantum protocol, we achieve a nominal three-node entanglement generation rate of around \SI{10}{\per\second}.


\section{Network Architecture and Entanglement Protocol}

\noindent Our network consists of three nodes, each containing one atom as memory qubit. The network nodes are located in three different laboratories, with the central one (Lab C) connected to the two remote nodes (Labs A and B) via tens of metres of optical fibres. An electro-optical switch at the central node's output allows for redirecting paths on a nanosecond timescale. The central node sequentially emits two photons, both being entangled with the atom. One photon is sent to each of the two remote nodes to generate pairwise entanglement between the two matter qubits, resulting in a three-partite entangled state.

\begin{figure}
    \centering
    \includegraphics[width=\linewidth]{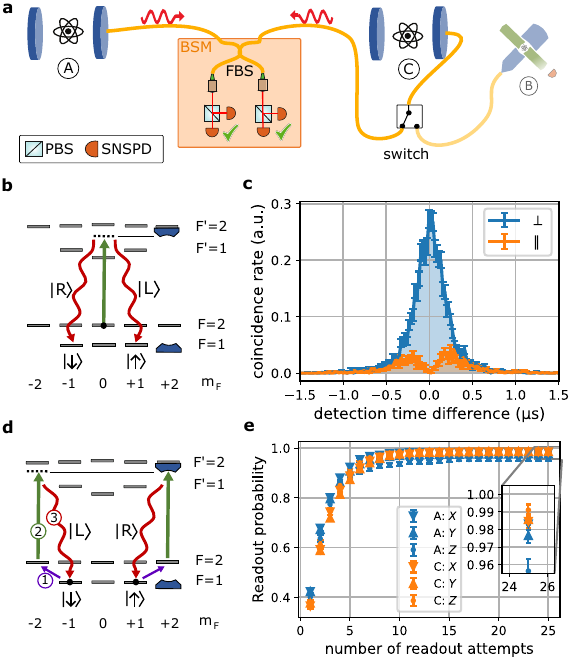}
    \caption{\textbf{Link characterization between Labs A and C}. \textbf{a}, At both labs, atom–photon entanglement is generated via the scheme depicted in \textbf{b}. In each lab, a vSTIRAP control pulse (green arrow) initiates the emission of a photon (red arrow) whose polarisation is entangled with the atomic state. Here, the switch after the cavity in Lab C routes the photon to the BSM setup composed of linear optics, i.e. fibre beam splitter (FBS) and polarizing beam splitters (PBS), and superconducting nanowire single-photon detectors (SNSPD). \textbf{c}, The resulting Hong–Ou–Mandel interference pattern for orthogonal ($\perp$) and parallel ($\parallel$) polarisation. The atoms in both labs are read out using the scheme shown in \textbf{d}: a Raman pulse (step 1) transfers the atom from the $\ket{F=1, m_F=\pm1}$ to the $\ket{F=2, m_F=\pm2}$ states. Excitation from these states (step 2) then produces a second photon (step 3) whose polarisation determines the atomic state. \textbf{e}, Cycling this readout scheme increases the readout probability.}
    \label{fig:link_1}
\end{figure}

The first link between Labs A and C is depicted in \Fig \ref{fig:link_1}a. The two nodes are identical setups consisting of a single $^{87}$Rb atom coupled to a macroscopic, high-finesse, optical Fabry-P\'{e}rot resonator. Each atom can emit a photon into the cavity mode, entangled with the atomic state via a vacuum-stimulated Raman adiabatic passage (vSTIRAP) (\Fig \ref{fig:link_1}b). This process allows for full control over the temporal modes of the photons through the intensity and phase profiles of the control lasers~\cite{MorinPRL2019}. The photonic qubits are defined by their polarization, with basis states $\ket{R}$ and $\ket{L}$ corresponding to right- and left-circular polarisation, respectively. The memory qubits are represented by $\ket{\uparrow_{A,C}} \equiv \ket{F=1, m_F=+1}$ and $\ket{\downarrow_{A,C}} \equiv \ket{F=1, m_F=-1}$ of the $5^2S_{1/2}$ ground state. Entanglement of the two atoms is realized via entanglement swapping by applying a BSM to the two photonic qubits from nodes A and C. The BSM is implemented with linear optics and heralds the preparation of the $\ket{\Psi^{\pm}}$ Bell-states. An important figure of merit for this particular scheme is the indistinguishability of the photons from each node. When properly matched, we observe a Hong-Ou-Mandel (HOM) visibility of up to \SI{80}{\percent} (\Fig \ref{fig:link_1}c) without temporal filtering.

Readout of the two memory qubits is based on a polarisation measurement of two single photons, each emitted by one atom, as shown in \Fig \ref{fig:link_1}d. The probability of obtaining a measurement outcome is given by the overall source-to-detection efficiency of each photon, which is approximately \SI{40}{\percent} in our case. However, this process can be repeated multiple times. This drastically increases the readout probability, which quickly converges to near-unity efficiency with the number of repetitions (see \Fig \ref{fig:link_1}e). To fully characterize the atomic states we measure the states of the atoms in different bases. To set the $X$ and $Y$ bases, a Raman $\pi$/2 pulse between $\ket{\uparrow_{A,C}} \leftrightarrow \ket{\downarrow_{A,C}}$ with an appropriately chosen phase is applied directly before the readout procedure.

\begin{figure}
    \centering
    \includegraphics[width=\linewidth]{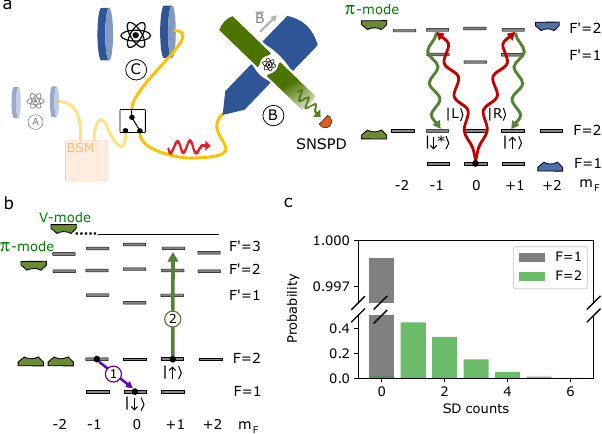}
    
    \caption{\textbf{Link characterization between Labs C and B.} \textbf{a}, Lab C generates an atom-photon entangled state via the emission process shown in \Fig \ref{fig:link_1}d. The photon is directed by an optical switch to Lab B, where it is stored in the heralded memory. As the photon (red arrows) interacts with the atom via one cavity, a second photon (green arrow) is emitted in the $\pi$-mode of the second cavity and is directed to a SNSPD. Its detection heralds the successful mapping of atom-photon into atom-atom entanglement. \textbf{b}, To readout the atom of node B, a microwave $\pi$-pulse (step 1) transfers $\ket{\downarrow^{\ast}_{B}}$ to $\ket{\downarrow_{B}}$. Afterwards, (step 2) fluorescence state detection of the remaining $\ket{F=2}$ population is performed on the cycling transition $\ket{F=2} \leftrightarrow \ket{F'=3}$, which scatters photons efficiently into both herald cavity eigenmodes. Based on the orientation of their electric field to the magnetic field $\vec{B}$, the cavity eigenmodes are named $\pi$ (parallel to $\vec{B}$) and $V$ (orthogonal to $\vec{B}$). As soon as one photon is detected by the SNSPD, the resonant light is switched off after a feedback delay of \SI{500}{\nano\second}, which explains the average photon number of 1.8 $>$ 1 for $\ket{F=2}$, shown in the histogram in \textbf{c}.}
    \label{fig:CavX}
\end{figure}

To establish the entanglement between the two atoms in Labs B and C, the central node emits an entangled photon guided to node B. This third node differs from the other two nodes, as it consists of a $^{87}$Rb atom simultaneously coupled to two crossed fibre-based high-finesse Fabry-P\'{e}rot micro-resonators. Node B features the capability to store an incoming photonic qubit in the atomic state via one cavity while emitting a herald photon into the second (birefringent) cavity (see \Fig \ref{fig:CavX}a)~\cite{Brekenfeld2020}. The splitting of the polarisation eigenmodes of the herald cavity is essential for selecting a specific atomic decay channel from several available paths. Here, the memory qubits are represented by $\ket{\uparrow_{B}} \equiv \ket{F=2, m_F=+1}$ and $\ket{\downarrow^{\ast}_{B}} \equiv \ket{F=2, m_F=-1}$ of the $5^2S_{1/2}$ ground state. Immediately following the memory process, a microwave (MW) $\pi$ pulse remaps the atomic qubit to $\ket{\uparrow_{B}}$ and $\ket{\downarrow_{B}} \equiv \ket{F=1, m_F=0}$.

Readout of node B is achieved using state detection (SD) via cavity-enhanced fluorescence on an atomic cycling transition. The photons emitted during this process are efficiently collected by both eigenmodes of the herald cavity (\Fig \ref{fig:CavX}b). In order to minimize heating of the atom, the resonant excitation beam is switched off as soon as one photon is detected. \Fig \ref{fig:CavX}c shows a histogram of the SD counts for the states $\ket{F=1}$ (dark) and $\ket{F=2}$ (bright). Compared to nodes A and C, this readout process always yields an outcome with a discrimination fidelity of \SI{99(1)}{\percent}. Similarly to the other nodes, the $X$ and $Y$ readout basis is set by applying a MW $\pi$/2 pulse between $\ket{\downarrow_{B}} \leftrightarrow \ket{\uparrow_{B}}$ with an appropriately chosen phase.

\section{Two-Nodes Entanglement}

\begin{figure}[b]
    \centering
    \includegraphics[width=\linewidth]{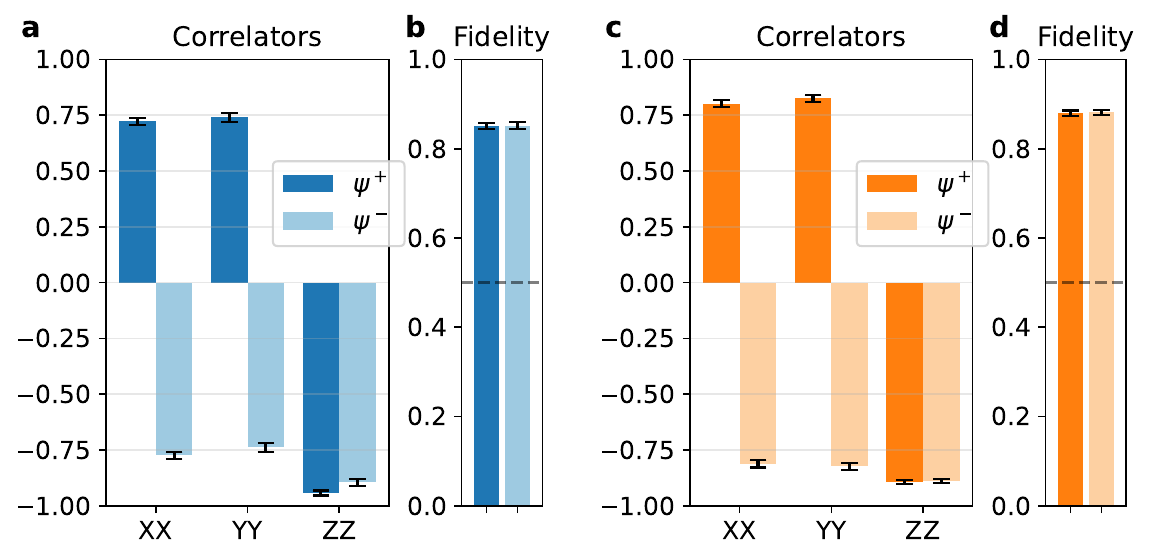}
    \caption{\textbf{Characterization of the two elementary links of the network.} \textbf{a}, Two-qubit correlators for the Bell-states generated between nodes A and C.
    \textbf{b}, Corresponding fidelity $F_{\psi^\pm} = (1\pm\mean{XX}\pm\mean{YY}-\mean{ZZ}{})/4$. 
    \textbf{c-d}, same as \textbf{a,b} but for the link connecting nodes C and B. (Error bars correspond to one standard deviation.)}
    \label{fig:2-node_entanglement}
\end{figure}

\noindent The entanglement fidelity of each link is characterized independently by measuring the $XX$, $YY$ and $ZZ$ correlators of the two memories involved. For the link A-C, we find that the two Bell-states, which can be distinguished by the BSM detection pattern, have an average fidelity of \SI{80.5(5)}{\percent} (\Fig \ref{fig:2-node_entanglement} a). This is primarily limited by the HOM visibility. However, the fidelity can be increased by temporally filtering the BSM photons, which sacrifices efficiency. Unless otherwise stated, we truncate \SI{17.5}{\percent} of the tail of the photons and obtain an average fidelity of \SI{85.2(5)}{\percent}. For the second link between nodes C and B, the central node is prepared in a superposition state $\ket{\uparrow_C}\pm\ket{\downarrow_C}$ before sending the entangled photon (emitted via \Fig \ref{fig:link_1}d) to node B, resulting in the preparation of a Bell-state upon detection of the herald photon (\Fig \ref{fig:2-node_entanglement}c). For this link the average fidelity of the Bell-states is \SI{88.0(4)}{\percent}.

\section{Three-Nodes Entanglement}

\noindent The entanglement of the three nodes is achieved by having the central node emitting successively two photons forming an atom-photon-photon GHZ state~\cite{Thomas2022}. Each photon is sent to one node (A, B) to entangle each link using the processes described in the previous paragraphs(see Methods for details). We use the qubit definition discussed earlier but apply a bit flip on atom C such that the generated state formally matches the GHZ state in literature
\begin{equation}
    \frac{\ket{\uparrow_A\uparrow_B\uparrow_C}+\ket{\downarrow_A\downarrow_B\downarrow_C}}{\sqrt{2}}.
\end{equation}


\begin{figure*}
    \centering
    \includegraphics[width=\linewidth]{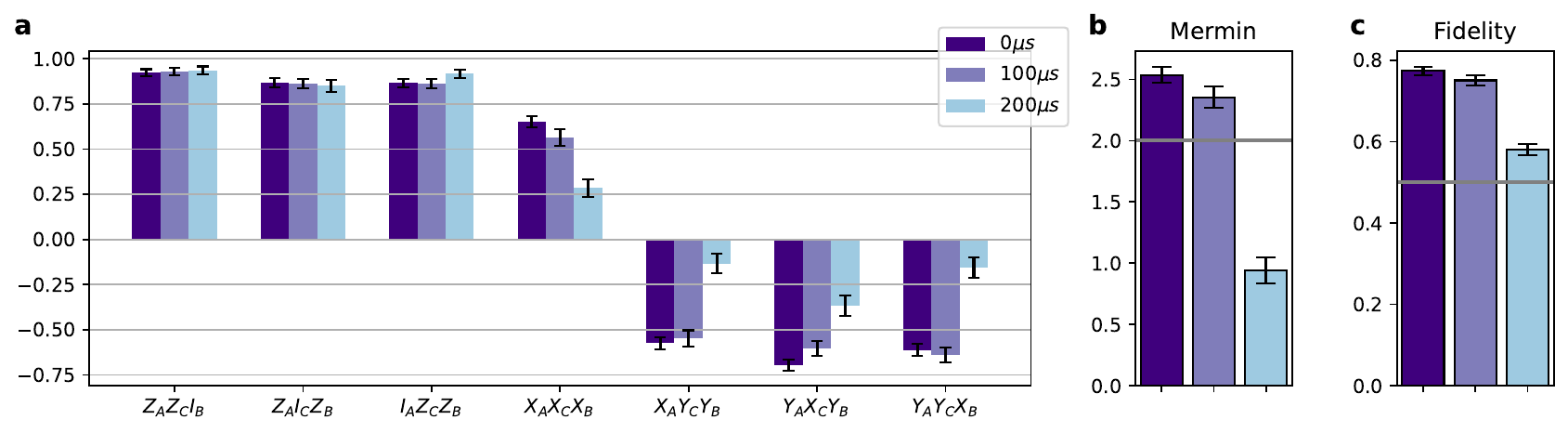}
    \caption{\textbf{Three-node GHZ-state characterization and storage.} 
    \textbf{a},  Measured three-node correlations after storage times of 0, 100 and \SI{200}{\micro\second}. The decay of the correlators involving only $X$ and $Y$ measurements, while the population correlators ($Z$ measurements) remain unchanged, indicates that dephasing is the dominant decoherence mechanism. 
    \textbf{b}, Corresponding Mermin parameter. The measured values exceed the classical threshold of 2 for storage times up to \SI{100}{\micro\second}. 
    \textbf{c}, GHZ-state fidelity. Owing to the contribution of the population terms, the fidelity is less sensitive to dephasing and remains above the threshold of 0.5 for genuine tripartite entanglement even after \SI{200}{\micro\second} of storage. (Error bars correspond to one standard deviation.)
    }
    \label{fig:5-node_results}
\end{figure*}

In order to characterize the three-node entanglement, we measure the 7 correlators $ZZI$, $ZIZ$, $IZZ$, $XXX$, $XYY$, $YXY$, $YYX$ (see Methods for the basis settings). In total, we detected around 8000 GHZ states out of 9400 heralded events within a wall-clock time of 10.5 hours, with at least 300 events observed per measurement setting. To show that the state can be stored in the memory qubits, we measure the correlators for different storage times as shown in \Fig \ref{fig:5-node_results}a. Decoherence, which only affects the $X$ and $Y$ measurements, is primarily caused by magnetic field fluctuations that alter the phase of the superposition. Thus, the $Z$ correlators remain essentially unchanged as the storage time increases, while the $X$ and $Y$ correlators decrease significantly.

An important benchmark for multipartite entanglement is the Mermin parameter~\cite{Mermin1990}, representing a non-locality test. For three particles, it is given by $M=\mean{XXX-XYY-YXY-YYX}$. To assess non-locality, the measured Mermin parameter should be greater than 2. However, in our experiment, nodes A and C lack deterministic readout, opening a detection loophole~\cite{Brunner2014}. The standard local-hidden-variable bound of 2 therefore no longer applies, and a corrected threshold that accounts for the finite readout efficiencies must be used instead~\cite{Larsson1998}. For the \SI{0}{\micro\second} dataset, the efficiencies in the $X$ and $Y$ bases are identical within uncertainties (see \Fig~\ref{fig:link_1}e). We therefore assume a basis-independent efficiency and define conservative bounds by taking the extremal values that are compatible with the measured efficiencies and their one-standard-deviation uncertainties. Using 25 readout attempts, the bounds are $0.969 \leq \eta_A \leq 0.980$ and $0.979 \leq \eta_C \leq 0.985$, resulting in a new threshold of $2.07 \leq M_{\mathrm{LHV}} \leq 2.11$. Since the measured Mermin parameter is $M=\SI{2.54(6)}{}$ at \SI{0}{\micro\second}, a clear violation remains even when considering the detection loophole. A natural extension would be to perform a Svetlichny test to prove three-partite non-locality~\cite{Brunner2014}. Since the Mermin parameter only involves $X$ and $Y$ measurements, it is only linked to the coherences of the state. This makes it extremely sensitive to the decoherence of the superposition. As shown in \Fig~\ref{fig:5-node_results}b, for \SI{200}{\micro\second}, the effect of decoherence caused by the combined finite coherence times of the individual memories (see Methods) is clearly visible.

Finally, using all correlators, we compute the fidelity $F = (1+\mean{ZZI}+\mean{ZIZ}+\mean{IZZ}+\mean{XXX}-\mean{XYY}-\mean{YXY}-\mean{YYX})/8$~\cite{Guehne2009}. It guarantees genuine multi-particle entanglement if the fidelity is above 0.5. As shown in \Fig~\ref{fig:5-node_results}c, all three coherence time measurements exceed this threshold, reaching values of $F = \SI{77(1)}{\percent}$ for \SI{0}{\micro\second}. Since the fidelity includes the population terms ($Z$ measurements), it is less impacted by the decoherence compared to the Mermin parameter.

\section{Entanglement Efficiency}

\noindent Another important figure of merit is the entanglement efficiency and rate. We measured a total three-partite entanglement efficiency of \SI{0.16(1)}{\percent}, where the uncertainty reflects the fluctuations over two days of measurement. This number comprises several contributions, which we will discuss in the following. First, we estimate the success probability for entanglement over the A-C link from the coincident detection of two photons and a factor of 0.5 from the BSM realized with linear optics. The typical photon-source-to-detection efficiency of the process in \Fig \ref{fig:link_1}b is around \SI{30}{\percent}, which includes the preparation of the state $\ket{F=2, m_F=0}$. Considering \SI{83}{\percent} temporal filtering efficiency we have a good agreement with the measured efficiency of $\eta_\text{AC}\simeq \SI{4}{\percent}$. Second, for the C-B link, the entanglement success probability is governed by the photon-source-to-memory transfer efficiency and the heralding efficiency of node B, which we estimate to be \SI{40}{\percent} and \SI{10}{\percent}, respectively. This coincides with the measured efficiency of $\eta_\text{CB}\simeq \SI{4}{\percent}$. Therefore, the combined success probability of entangling both links is $\eta_\text{AC}\ \eta_\text{CB}\simeq \SI{0.16}{\percent}$ which matches the measured value. To determine the nominal entanglement rate, we must consider the time, $\tau_\text{ent}=\SI{140}{\micro\second}$ (see Methods), required to generate three-partite entanglement. Thus, the nominal entanglement rate is $R_\text{nom}=\eta_\text{AC}\ \eta_\text{CB} / \tau_\text{ent} \simeq \SI{11}{\per\second}$. Technical limitations are discussed in the Methods section. 

With this in mind, it is instructive to consider a scenario in which the central node C emits a third photon (via \Fig \ref{fig:link_1}d within \SI{50}{\micro\second}) to entangle a third link with a fourth node, similar to nodes A and B ($\eta_\text{CD}\simeq \SI{4}{\percent}$). The corresponding nominal rate for such \textit{four}-node network would be $\eta_\text{AC}\ \eta_\text{CB}\ \eta_\text{CD} / (\tau_\text{ent}+\SI{50}{\micro\second}) \simeq \SI{20}{\per\minute}$. Considering the reduction in rate due to the technical limitations (see Methods), the achievable rate remains competitive with the rate of all previously realised \textit{three}-node networks~\cite{Jing2019, Pompili2021}. This demonstrates that efficient entanglement generation is a powerful lever and, in fact, a necessary condition for practical entanglement distribution to multiple nodes.

The estimated rate can be enhanced even further, especially for long-distance links for which the communication time is not negligible, by operating each node with multiple atoms rather than one~\cite{Avis2023,Hartung2024,Krutyanskiy2024}. With a repeat-until-success strategy, each link could then be entangled at a higher rate by means of a multiplexing protocol. The temporal sequence of the protocol, together with the timestamps of the recorded heralding photons, can be used to determine which atoms are entangled. Such multiplexing has the advantage that different atoms can be entangled independently from each other.

\section{Conclusion and outlook}

\noindent In conclusion, we distributed a long-lived GHZ state across three quantum nodes housed in three separate laboratories with an efficiency of \SI{0.16}{\percent} and a nominal rate of around \SI{10}{\per\second}, which is unprecedented. The generated state reaches a fidelity of \SI{77(1)}{\percent}, and the measured correlations violate Mermin's inequality. The state remains coherent for at least \SI{200}{\micro\second} of storage time. Spin-echo, dynamical-decoupling or optical-transfer techniques might boost this lifetime to tens or even hundreds of ms~\cite{Koerber2018}.

Looking into the future, we envision a quantum network in which a central node emits photons with a tailor-made entanglement topology~\cite{Hein2006}, of which the GHZ state realised here is just one example. These photons are individually steered to different end nodes which store the entanglement as a versatile resource. Appropriate local measurements on a subset of nodes would leave the remaining nodes in a desired entangled state independently of the fibre network topology.~\cite{Briegel1998}. For example, point-to-point entanglement could be established between any chosen pair of nodes, effectively treating the measured nodes as repeater nodes that are removed from the network~\cite{Epping2016}. Further fascinating possibilities open up when, for instance, the two qubit states are encoded as suitable GHZ states, as the corresponding superpositions are predicted to exhibit inherent stability against noise and decoherence in the network~\cite{Froewis2011}. With all this in mind, the efficient, heralded entanglement achieved by means of an optical cavity provides the building block for a reconfigurable network, and ultimately for all the fascinating applications of a quantum internet~\cite{Wehner2018, Awschalom2021}.

\bibliographystyle{plain}

\let\oldaddcontentsline\addcontentsline
\renewcommand{\addcontentsline}[3]{}

\section*{Methods}
\renewcommand{\thefigure}{\textbf{\arabic{figure}}}
\renewcommand{\theHfigure}{methods\arabic{figure}}
\renewcommand{\figurename}{\textbf{Extended Data \Fig}}
\setcounter{figure}{0}
\renewcommand{\tablename}{\textbf{Extended Data Table}}
\renewcommand{\thetable}{\arabic{table}}
\setcounter{table}{0}

\section*{Entanglement generation rate}
\noindent Our nominal rate of \SI{11}{\per\second} is defined as the product of the success probability and the attempt frequency:  $R_\text{nom}=\eta_\text{AC}\ \eta_\text{CB} / \tau_\text{ent}$. Our raw rate is defined as the number of successfully heralded GHZ states divided by the total acquisition time: $R_\text{raw}=\SI{15}{\per\minute}$ (\SI{0.25}{\per\second}). The factor of around 50 between the two rates stems from various technical limitations and implementation choices.

\paragraph{Atom availability.}
Depending on the node and experimental conditions, atoms remain trapped for durations ranging from several seconds to approximately one minute. Reloading an atom typically takes one or two seconds. During the measurements reported here, the availability of the atoms was approximately \SI{55}{\percent}, \SI{55}{\percent}, and \SI{80}{\percent} for nodes A, B, and C, respectively. Since all three nodes must contain an atom simultaneously, the combined availability is $A_\text{atoms}\approx \SI{25}{\percent}$.

\paragraph{Cooling and imaging.}
Each experimental repetition consists of the quantum protocol, followed by a cooling and imaging stage. The fluorescence collected during the cooling phase is used to image the atomic position and ensure optimal coupling to the cavity mode. Typically, $D_\text{imaging}\approx \SI{50}{\percent}$ of the total duration of one experimental repetition is dedicated to cooling. This ratio is primarily determined by the need for a sufficient signal-to-noise ratio for the imaging, rather than by the cooling dynamics, which occur on a significantly shorter timescale. Therefore, a gated imaging system would allow for a more favourable ratio.

\paragraph{Readout.}
In our implementation, the readout is performed systematically to preserve any potentially useful data for future analysis, regardless of the BSM outcome or the heralding event from the node B. The readout takes up to 1.4 ms for 25 attempts, which is ten times longer than the entanglement-generation sequence itself, resulting in a duty cycle overhead of $D_\text{readout}\approx \SI{17}{\percent}$. Consequently, the average repetition rate is significantly impacted by the readout stage. This overhead could be avoided by implementing a dynamically changing sequence length in the real-time analysis of the Field Programmable Gate Array control system, which would restart as soon as a three-partite GHZ state was not heralded.

In summary, the raw rate is given by $R_\text{raw} = R_\text{nom}\, A_\text{atoms}\, D_\text{imaging}\, D_\text{readout}$. These three contributions reduce the rate by approximately a factor of 50. The atom readout contribution is obviously a choice rather than a technical limitation. The imaging and atom trapping contributions can be strongly reduced, but this requires more involved experimental efforts.

\section*{Coherence time}
\begin{figure}[h]
    \centering\includegraphics[width=\linewidth]{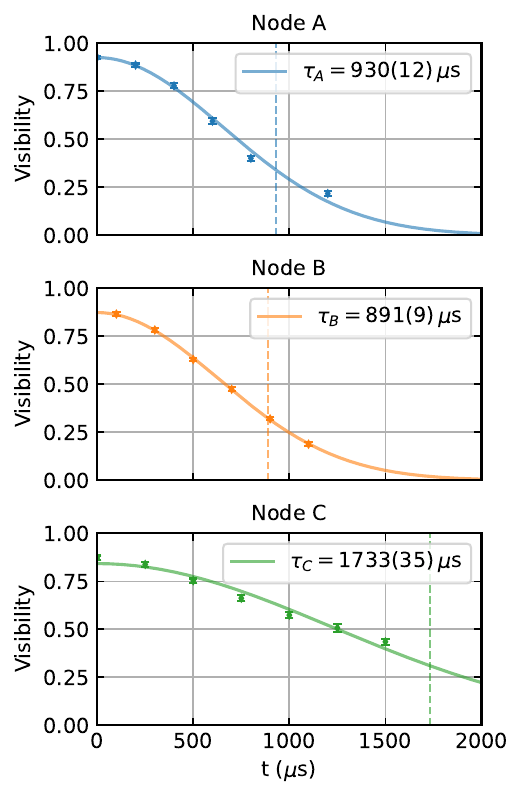}
    \caption{\textbf{Coherence times of the individual nodes}. In each lab, the atomic qubit was initialised in a superposition state $\ket{\uparrow_i}+\ket{\downarrow_i}$, where $i \in \{A,B,C\}$ at $t = \SI{0}{\micro\second}$. After time $t$ the atomic qubit is read out and the visibility of the superposition state is measured. A Gaussian fit is applied to the visibility values. The 1/e coherence time (dashed line) of the labs is given by: $\tau_A = \SI{930(12)}{\micro\second}, \tau_B = \SI{891(9)}{\micro\second}, \tau_C = \SI{1733(35)}{\micro\second}$.}
    \label{fig:coherence}
\end{figure}

\noindent To measure the coherence times in each laboratory, we initialise the atomic qubit in the superposition state $\ket{\uparrow_i}+\ket{\downarrow_i}$, where $i \in \{A,B,C\}$. After a period of time $t$, we read out the atomic state in the $X$ basis, as described in the main text, and measure the visibility of the superposition state. In each lab, the atomic qubit is sensitive to the magnetic field because it is encoded in different Zeeman sublevels. Magnetic field fluctuations cause a Gaussian decay. The 1/e coherence times in the individual labs are given by: $\tau_A = \SI{930(12)}{\micro\second}, \tau_B = \SI{891(9)}{\micro\second}, \tau_C = \SI{1733(35)}{\micro\second}$ (see Extended Data \Fig \ref{fig:coherence}). 

\section*{Sequence}

\begin{figure*}
    \centering
    \includegraphics[width=\linewidth]{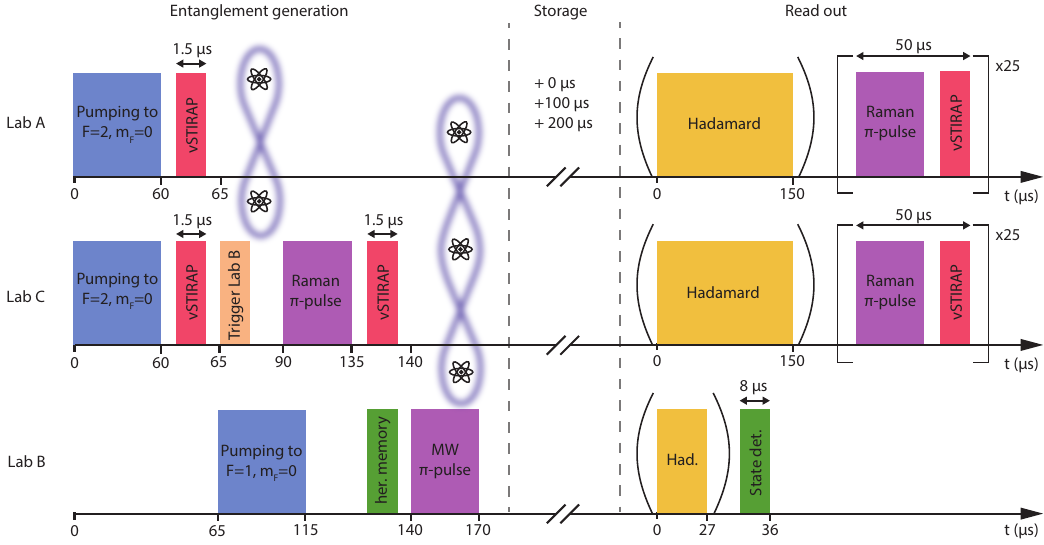}
    \caption{\textbf{Experimental sequence for GHZ states generation.} This sketch illustrates the pulse and trigger sequence used to generate GHZ states. The time axis is not to scale. The Hadamard gates are represented in parentheses because they are applied only when measuring the qubits in the $X$ and $Y$ bases.
    }
    \label{fig:sequence}
\end{figure*}
\noindent The experimental sequence is shown in Extended Data \Fig \ref{fig:sequence}. Labs A and C are synchronised. In both laboratories, the atom is pumped to the $\ket{F=2,m_F=0}$ state. Each atom generates an entangled photon using the vSTIRAP process, as shown in \Fig \ref{fig:link_1}b. Both photons are guided to the BSM setup, where a BSM is performed to project the two atoms onto an entangled state. An electrical signal send to Lab B triggers the preparation of the atom in the $\ket{F=1,m_F=0}$ state. Meanwhile, two Raman $\pi$ pulses are applied to the atom in Lab C (see \Fig \ref{fig:link_1}d step 1). A further vSTIRAP pulse (step 2) generates an additional photon which is entangled with the two atoms in Labs A and C. To prevent phase evolution of the entangled GHZ state during the emission process of this photon, the frequency difference between its polarisation components is compensated by \SI{400}{\kilo\hertz}. Further information can be found in the frequency compensation section below. This photon is then guided to Lab B, where a heralded memory procedure is performed (see \Fig \ref{fig:CavX}a). Detection of the herald photon indicates the successful entanglement of all three particles. This occurs within \SI{140}{\micro\second}. Immediately afterwards, a microwave $\pi$ pulse is applied in Lab B, to remap the qubit from $\{\ket{\downarrow^\ast_B},\ket{\uparrow_B}\}$ to $\{\ket{\downarrow_B},\ket{\uparrow_B}\}$ (see \Fig \ref{fig:CavX}d step 1). The three-partite entangled state is stored for up to \SI{200}{\micro\second}. To perform measurements in the $X$ and $Y$ basis, a Hadamard gate is applied to all the three atoms. Further information about the Hadamard gate can be found in the Basis settings section. To read out the atoms in Labs A and C, the atomic qubit is mapped onto a photon whose polarisation is measured (see \Fig \ref{fig:link_1}d). The photons from the two nodes are generated sequentially, \SI{3}{\micro\second} apart, as both nodes use the BSM setup for photon detection. This probabilistic process takes \SI{50}{\micro\second} and is repeated up to 25 times. Readout in Lab B is performed using cavity-enhanced fluorescence state detection. This process is deterministic and takes only \SI{8}{\micro\second}.

\section*{Basis settings/calibrations}
\noindent To ensure meaningful quantum correlations across the network, it is essential to ensure compensated quantum channels and a shared reference frame, i.e. the same basis alignment for all the nodes in the network.

We first focus on the compensation of the quantum channels. To compensate the quantum channel between Labs A and C, we consider the three components, the optical fibre between Lab A and the BSM setup, the optical fibre between Lab C and the BSM setup, and the BSM setup itself. To align the polarisation basis of the BSM setup, we use wave-plates at each output of the fibre beam splitter. This ensures that any arriving polarization qubit is transformed in the very same way, regardless of the BSM detection arm. This establishes a polarization identity transformation between the two polarization analysis modules of the BSM setup. The optical fibres between Labs A and C and the BSM setup are compensated with wave-plates at the output of each cavity, ensuring that a right-circularly polarised input photon is always reflected by the PBSs. This guarantees that Labs A and C share the same photonic $Z$ measurement basis. In fact, the photonic $X$ and $Y$ measurement bases are most likely different. However, this can be compensated for by rotating the atomic qubits appropriately (see the following paragraph). Similarly, the quantum channel between Labs C and B is compensated with wave-plates at the input of node B, ensuring that an emitted right-circularly polarised photon at Lab C is mapped to a right-circularly polarised photon arriving at Lab B. This guarantees same photonic $Z$ measurement basis between the two Labs. However, there may still be some polarisation rotation for linearly polarised input photons. This can be again compensated for by applying an appropriate rotation to the atomic qubit.

Measuring meaningful quantum correlations between all three atoms in the $Z$ basis is already possible by compensating the two quantum channels as mentioned above. In order to read out the atoms in the $X$ and $Y$ bases, a Hadamard gate must be performed on each atom before the corresponding atomic qubit is measured. For Labs A and C, this is realised using a Raman $\pi$/2 pulse between $\ket{\uparrow_{A,C}} \leftrightarrow \ket{\downarrow_{A,C}}$. This involves three pulses: a Raman $\pi$ pulse between $\ket{\uparrow_{A,C}} \leftrightarrow \ket{F=2, m_F=0}$, followed by a Raman $\pi$/2 pulse between $ \ket{F=2, m_F=0} \leftrightarrow \ket{\downarrow_{A,C}}$, and then another Raman $\pi$ pulse between $\ket{\uparrow_{A,C}} \leftrightarrow \ket{F=2, m_F=0}$. The phase of the Hadamard gate in Labs A and C can be tuned by shifting the timing of all three, since the atomic qubits precess at a frequency of \SI{200}{\kilo\hertz} due to the applied magnetic field along the cavity axis (for Lab B it is along the qubit cavity). The Larmor frequency is given by \SI{100}{\kilo\hertz}. The Hadamard gate in Lab B is realised using a MW $\pi$/2 pulse between $\ket{\downarrow_{B}} \leftrightarrow \ket{\uparrow_{B}}$. Here, it is possible to change the phase of the $\pi$/2 pulse directly by changing the phase of the MW.
To align the atomic $X$ and $Y$ bases across the network, we follow this procedure. First, we set the Hadamard phases in Lab C; then in Lab A and finally in Lab B. To set the phases in Lab C, a three-particle GHZ state consisting of two photons and the atom is generated. The phase of the Hadamard gate for the $X$ basis is then set to maximise correlations. The $Y$ basis is set by changing the phase by \SI{90}{\degree}. This is achieved by delaying the three Raman pulses by $1/4 \times 1/\SI{200}{kHz} = \SI{1.25}{\micro\second}$ with respect to the timing obtained for the $X$ basis. Lab C still generates the same GHZ state in order to set the atomic detection basis in Lab A. This time, however, a BSM of the first photon is performed by overlapping it with a photon from Lab A, resulting in entanglement between the two atoms and the second photon from Lab C. The phases of the Hadamard gate in Lab A are set analogously to those in Lab C. For the $X$ basis the correlations are maximised, whereas for the $Y$ basis, the timing of the three Raman pulses is changed by \SI{1.25}{\micro\second}. Finally, the atomic detection basis in Lab B is set by generating atom-atom-photon entanglement between Labs C and B. Lab C generates the same GHZ state again, but this time the first photon is detected using the BSM setup and the second photon is sent to Lab B. To set the correct phase of the Hadamard gate for the $X$ basis, the timing of the trigger sent from Lab C to Lab B is scanned (see Extended Data \Fig \ref{fig:sequence}). This alters the time spent in the $\{\ket{\downarrow^\ast_B},\ket{\uparrow_B}\}$ qubit, which precesses at a frequency of \SI{200}{\kilo\hertz}. A phase is accumulated, which is set to maximise the correlations with the other two particles. The $Y$ basis is obtained by changing the phase of the MW by \SI{90}{\degree}.

With this approach, it is possible to calibrate the detection basis of all three atoms by performing single- or two-node experiments only. This is much quicker and more efficient than three-node calibrations.

\section*{Frequency Compensation}

\begin{figure}[htbp]
    \centering\includegraphics[width=\linewidth]{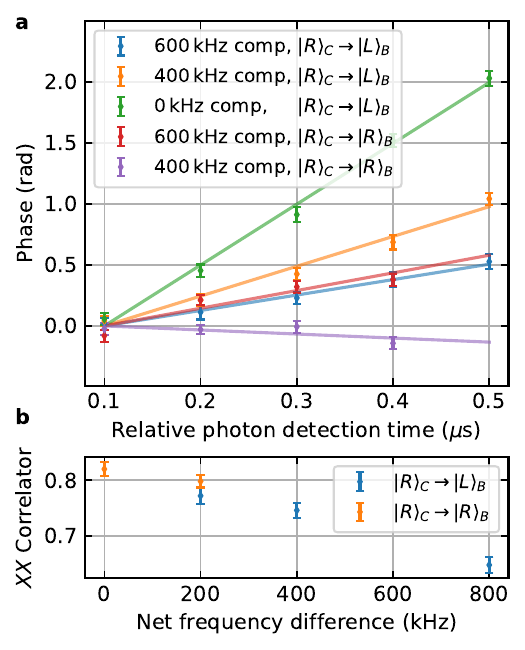}
    \caption{\textbf{Polarisation dependent frequency compensation.}. \textbf{a)}, The phase corresponding to the maximum correlator is plotted as a function of the time slice of a photon. To fully compensate for the phase evolution of the entangled state, a compensation of \SI{400}{\kilo\hertz} and a mapping from $\ket{R}_C$ to $\ket{R}_B$ are needed. This maximises the $XX$ correlators shown in \textbf{b)}.}
    \label{fig:frequency_compensation}
\end{figure}

\noindent In all three laboratories, a Larmor frequency of \SI{100}{\kilo\hertz} is applied, resulting in a frequency difference of \SI{200}{\kilo\hertz} between the $\{\ket{\downarrow_A}, \ket{\downarrow^\ast_B}, \ket{\downarrow_C}\}$ and the $\{\ket{\uparrow_A}, \ket{\uparrow_B}, \ket{\uparrow_C}\}$ states. The photons generated via the process described in \Fig \ref{fig:link_1}b also exhibit a frequency difference of \SI{200}{\kilo\hertz} between the two polarisation components, $\ket{R}$ and $\ket{L}$. Correctly compensating the quantum channel of the link between nodes A and C (see Basis Settings/Calibrations), results in a net frequency difference of 0. However, the second photon emitted by node C via the process described in \Fig \ref{fig:link_1}d, exhibits a frequency difference of \SI{600}{\kilo\hertz}. This photon is stored in the $\{\ket{\downarrow^\ast_B}, \ket{\uparrow_B}\}$ qubit, resulting in a net frequency difference of \SI{400}{\kilo\hertz}. If this frequency difference is not compensated for, the three-partite GHZ state will evolve by \SI{400}{\kilo\hertz} during the heralded memory process. This would decrease the correlators in the $X$ and $Y$ bases. To address this problem, we have employed an EOM at the output of the central node (not shown in \Fig \ref{fig:CavX}a), which ramps the phase of one polarisation component during the emission process of the second photon. This changes the frequency difference between the two polarisation components to \SI{200}{\kilo\hertz}, resulting in a net frequency difference of 0 and increased correlators in the $X$ and $Y$ bases (see Extended Data \Fig \ref{fig:frequency_compensation}). Note that compensating the quantum channel between nodes C and B in the opposite way (i.e. mapping $\ket{R}_C$ to $\ket{L}_B$ instead of $\ket{R}_C$ to $\ket{R}_B$; see Basis Settings/Calibrations) also results in a net frequency difference of \SI{400}{\kilo\hertz}.

\section*{Acknowledgments}
\noindent We thank Darius Haitsch for experimental assistance on the A-C link, and Mullai Sampangi Raman and Leart Zuka for FPGA programming. This work was funded by the Deutsche Forschungsgemeinschaft (German Research Foundation) under Germany’s Excellence Strategy – EXC-2111 – 390814868, by the German Federal Ministry of Research, Technology and Space (Bundesministerium für Forschung, Technik und Raumfahrt, BMFTR) through the project QR.N (16KIS2189), as well as by the Munich Quantum Valley lighthouse project NeQuS (Z.5-F5121.17.1/5/80) under the Hightech Agenda Bayern Plus of the Bavarian state government and by the European Union’s Horizon 2020 research and innovation programme via the project Quantum Internet Alliance (QIA, GA No. 820445).

During the writing of this manuscript, we became aware of related work in the group of C. Monroe at Duke University, USA.

\end{document}